\documentclass[eqsecnum,aps,tightenlines]{revtex4}
\usepackage{epsfig}

\begin{document}
\title{
Centrality Dependence of Thermal Parameters in 
Heavy-Ion Collisions at CERN-SPS}
\author{
J. Cleymans$^a$, B. Kampfer$^b$ and S. Wheaton$^a$,
}
\affiliation{
$^a$Department of Physics, University of Cape Town,\\
Rondebosch 7701, Cape Town, South Africa\\
$^b$Institut f\"ur Kern- und Hadronenphysik,\\ 
Forschungszentrum Rossendorf, PF 510119\\
D-01314 Dresden, Germany
}


\begin{abstract}
The centrality dependence of thermal parameters,
characterizing the hadron multiplicities, 
is determined phenomenologically
for lead-on-lead collisions at CERN-SPS for a beam energy of 158 AGeV.
The strangeness equilibration factor  shows a clear, approximately linear,
increase with increasing centrality, while the freeze-out
temperature and chemical potential
remain constant. \\[3mm]
\end{abstract}

\maketitle

It has been shown that the abundances of different hadronic 
species in the final state of relativistic heavy-ion collisions
can be well described by statistical-thermal models 
(cf.\ \cite{Redlich,Heppe,Becattini} and further references therein).
In such a way, the observed multiplicities of a large number of hadrons
can be reproduced by a small number of 
parameters \cite{Heppe,Becattini}, such as 
the temperature, 
the baryon chemical potential,
the volume, 
and a parameter \cite{Slotta} measuring the degree of
equilibration of strange particles.
Such a description can be justified for multiplicities
measured over the whole phase-space, since many dynamical effects 
cancel out in ratios of hadron yields; in particular, effects due
to flow can be shown to disappear if the freeze-out surface
is characterized by a single temperature and chemical potential
\cite{prc,Rischke}.

As a matter of fact we mention that electromagnetic signals,
i.e. real and virtual photons, observed in central heavy-ion
collisions at CERN  Super Proton Synchrotron (SPS) energies 
can also  be described astonishingly well
within the framework of a thermal fireball model \cite{e.m.signals}.
This is not necessarily a proof for 
(global) thermalization \cite{Rischke}, but shows that
the use of thermal models allows a very economical description
of a large set of observables by a few characteristic parameters.   

It is the subject of the present paper to pursue this idea
and to analyze the centrality dependence of the thermal
parameters describing hadron multiplicities.
This will provide further information 
about the effects of the size of the excited strongly interacting
system and help in the systematic understanding of the experimental data. 
We will show that the thermal model is able to
describe the available data for various centrality classes at one beam energy.

In order to have a sound basis for the application of the 
thermal model we rely as much as possible
on fully integrated particle multiplicities. For this reason we 
concentrate our efforts on the analysis of  results obtained by 
the NA49 collaboration \cite{Sikler} using 
centrality selected fixed-target Pb + Pb collisions
at a beam energy of 158 GeV per nucleon, which are here analyzed
within the framework of the thermal model \cite{Becattini}.
The data \cite{Sikler} consists of centrality
binned multiplicities of the following hadronic species:
$\bar{p}$, $ \langle \pi \rangle = (\pi^+ + \pi^-)/2$, 
$K^+$ and $K^-$,  as a function of the number of participants.
Following \cite{Sikler,Cooper}, these centrality classes
correspond to mean impact parameters (in units of fm)
2.2 (bin I),
4.3 (bin II),
6.0 (bin III),
7.4 (bin IV),
8.9 (bin V), and
10.7 (bin VI).
The mean number of participants is listed in Table 1.
Notice that the number of hadronic species at our disposal 
is quite restricted in comparison with other available data sets for
central collisions. We  investigate here the
general trend for changing centrality, relying on the 
data set given in \cite{Sikler}. It should be emphasized that 
we have not included in our analysis 
the multiplicity of protons. At larger impact parameters many
protons are spectators which should not be
included in the thermal model analysis.

For completeness we  briefly recall the basic features of
 the statistical-thermal
model used to extract the chemical freeze-out
parameters. A more detailed description can be 
found in \cite{Becattini,Satz}.
\begin{itemize}
\item 
An ideal gas of all hadrons and hadronic resonances 
listed by the particle data group \cite{PDB} was used.
\item 
The grand-canonical ensemble description was used throughout
with the additional parameter $\gamma_s$ 
(cf.\ \cite{Slotta})
included to account
for the incomplete equilibration in the strange sector.
\item 
The correct quantum statistics was used for each particle.
\item 
Resonances were described with
a Breit-Wigner distribution \cite{Becattini}
over an interval [$m$ - $\delta m$, $m$ + $2\Gamma$]
where  
$
\delta m = \mathrm{min}[m_{\mathrm{threshold}},2\Gamma].
$
The  resonance widths $\Gamma$ were
taken from \cite{PDB}. 
\item 
All resonances were allowed to decay  with branching ratios 
taken from \cite{PDB}.
Weak decays were excluded since the data in \cite{Sikler} were
corrected accordingly. 
\item 
No excluded-volume corrections (cf.\ \cite{Gorenstein}) 
were incorporated. This does not affect the 
intensive thermal parameters ($T$, $\mu_B$, etc...), however,
corrections to the densities will be called for, in particular
the baryon and the energy densities.
\end{itemize}

According to our proposition,
at chemical freeze-out the system is completely  specified by
its temperature $T$, effective
volume $V_{\mathrm{eff}}$, strangeness suppression
factor $\gamma_s$ and chemical potentials $\mu_B$,
$\mu_S$ and $\mu_Q$. The initial conditions
of the collision fix $\mu_S$ and $\mu_Q$ for a
given $\mu_B$; $\mu_S$ is fixed by the requirement
of strangeness conservation, while $\mu_Q$ is
determined by the constraint that the ratio of the total charge
to total baryon number of the system is conserved.
 
The chi-square minimization routine MINUIT
was used to fit the chemical freeze-out parameters using\newline
$
N_{i} = \sum_{j} Br(j\rightarrow i) \: n_j^{0(\mathrm{prim})} \:
V_{\mathrm{eff}}
$
where $Br(j\rightarrow i)$ is the branching ratio of the decay
of hadron species $j$ into species $i$, and $ n_j^{0(\mathrm{prim})}$
denotes the primary density.
The values of $T$, $\mu_B$, $\gamma_s$ and $V_{\rm eff}$ were used as
free fit parameters adjusted to the fully integrated yields
$\bar{p}$, $K^+$, $K^-$, $\langle \pi \rangle$ and $N_{part}$ from 
\cite{Sikler}.
The results obtained for each of the six centrality bins 
are listed in Table~1 and displayed in Figs.~1 - 4.
The number of participants, $N_{\rm part}$, is identified with the 
baryon number of the  hadron gas.

As seen in Table~1,
good agreement exists between the trends in the data and
our model results.
As the number of pions has a very small
experimental error, the minimization routine zooms
in on them and fixes the parameters so as to
reproduce this number as accurately as 
possible. 
On the contrary, the number of participants, $N_{part}$, has a large
error  and  the minimization routine 
gives  much less weight
to them, hence the agreement is not as good.

Confident that the thermal model reproduces the
available hadron multiplicities in each of the centrality bins
with sufficient accuracy, let us consider the trends of the thermal
model parameters. The chemical freeze-out temperature and the 
chemical potential stay fairly constant for the various centralities,
as seen in Figs.~1 and 2.

A comparison of $T$ and $\mu_B$ with the values obtained in the recent
analysis of \cite{Becattini} for the most central collisions
shows that they are  consistent with the values of the  
present analysis. 
The results from \cite{Becattini} are shown as a band in 
Figs.~1 and 2.
As proven in \cite{Becattini},
the inclusion or omission of certain hadron species can change 
considerably the extracted values of $T$ and $\mu_B$. We 
stress however that our analysis, due to the restricted available
data, focuses on the trends with changing centrality.
The temperature, $T$, and the baryon chemical potential, $\mu_B$,
do not show any noticeable  dependence on centrality.

There are two thermal parameters which exhibit a pronounced
dependence on  centrality: the strangeness equilibration
factor $\gamma_s$ increases 
approximately linearly and then  saturates with increasing centrality,
as shown in Fig.~3; similarly, 
the radius $R$ of the hadron gas,
which can be readily compared to the radius of a static Pb 
nucleus increases approximately linearly with 
the number of participants, $N_{part}$, as seen in Fig~4.

Fig.~5 
shows the centrality dependence of various thermodynamic state variables,
such as the energy per hadron $\langle E \rangle / \langle N \rangle$,
the energy density, the baryon density, and the entropy per baryon
$S/B$. They all remain fairly independent of the centrality. 
In contrast, the Wroblewski factor \cite{wroblewski}, defined as
\begin{equation}
\lambda_s = {2\left<s\bar{s}\right>
\over \left<u\bar{u}\right> + \left<d\bar{d}\right>}
\end{equation}
increases nearly linearly with increasing centrality.

The predictions of other
hadron multiplicities, such as $K^0_s$, $\phi$, $\Xi^{\pm}$, and $\Lambda$,
$\bar\Lambda$ can serve as a further test of the thermal model.
These are not yet at our disposal in such a centrality binned way as
the yields of $\pi^{\pm}$, $K^{\pm}$, and $\bar p$ but, may become accessible
in further data analyses. We therefore present our prediction
in Table~\ref{predictions}.

In summary,
the analysis of the thermal parameters, describing the 
 integrated yields of $\pi^{\pm}$, $K^{\pm}$, and $\bar p$
as obtained by the NA49 experiment \cite{Sikler}, 
shows that  
the radius of the fireball increases linearly
with increasing centrality. Also,
the strangeness parameter $\gamma_s$ increases,
i.e., strange particle multiplicities approach  chemical equilibrium.
In contrast, the temperature 
and the baryon chemical potential 
do not change  with centrality.
No in-medium modifications are needed to describe the above quoted
hadron yields. 
\section*{Acknowledgments}
We acknowledge helpful correspondence and discussions 
with P. Jacobs, D. R\"ohrich, P. Seyboth, F. Sikler, and R. Stock 
on the NA49 data.
%

%

\begin{table} 
\begin{center}
\begin{tabular}{|c|c|c|c|c|c|c|}\hline
\multicolumn{1}{|c|}{Particle} &
\multicolumn{2}{|c|}{BIN I}  & 
\multicolumn{2}{|c|}{BIN II} &
\multicolumn{2}{|c|}{BIN III} \\ \cline{2-7}
\multicolumn{1}{|c|}{Species} &
\multicolumn{1}{|c|}{Data} &
\multicolumn{1}{|c|}{Model} & 
\multicolumn{1}{|c|}{Data} &
\multicolumn{1}{|c|}{Model} &
\multicolumn{1}{|c|}{Data} &
\multicolumn{1}{|c|}{Model}\\ \hline
$\langle \pi \rangle$  & 598   & 598   & 499.5 & 499.4 & 377 & 377 \\
$K^+$                  & 96.3  & 93.7  & 80.3  & 78.9  & 54  & 53.0 \\
$K^-$                  & 52.6  & 53.9  & 45.1  & 45.7  & 29.7& 30.1 \\
$\bar{p}$              & 10.4  & 10.4  & 8.6   & 8.6   & 6.94 & 6.93\\
$N_{\rm part}$         & 362   & 418   & 304   & 343   & 241 & 273 \\\hline\hline
\multicolumn{1}{|c|}{Particle} &
\multicolumn{2}{|c|}{BIN IV} &  
\multicolumn{2}{|c|}{BIN V} &
\multicolumn{2}{|c|}{BIN VI} \\ \cline{2-7}
\multicolumn{1}{|c|}{Species} &
\multicolumn{1}{|c|}{Data} &
\multicolumn{1}{|c|}{Model} & 
\multicolumn{1}{|c|}{Data} &
\multicolumn{1}{|c|}{Model} &
\multicolumn{1}{|c|}{Data} &
\multicolumn{1}{|c|}{Model}\\ \hline
$\langle \pi \rangle$ & 279  & 279  & 182.4  & 182.4 & 102 & 102 \\
$K^+$                 & 35.5 & 35.2 & 20.5 & 20.5& 10.3  & 9.8 \\
$K^-$                 & 20.2 & 20.3 & 11.8 & 11.8& 5.5   & 5.6 \\
$\bar{p}$             & 5.25 & 5.24 & 3.7  & 3.7 & 1.7   & 1.7 \\
$N_{\rm part}$        & 188  & 199  & 130  & 131 & 72 & 75 \\\hline
\end{tabular}
\end{center}
\caption{Comparison of experimental hadron yields with the
results of the thermal model with parameters as displayed
in Figs.~1 - 4.
}
\label{postdictions}
\end{table} 

\begin{table} 
\begin{center}
\begin{tabular}{|c|c|c|c|c|c|c|}\hline
Particle &BIN I  & BIN II &BIN III &BIN IV &  BIN V &BIN VI \\ 
{Species} & &  & & & &\\ \hline
$\Lambda$         & 50.5 & 42.2 & 29.4  & 19.4 &  11.4 & 5.5 \\
$\bar\Lambda$     & 5.97 & 5.00 & 3.54 & 2.34  & 1.45 & 0.56\\
$\Xi^-$           & 3.66 & 3.11 & 1.90 & 1.13  & 0.58 & 0.24 \\
$\overline{\Xi^-}$& 0.78 & 0.66 & 0.43 & 0.25  & 0.14 & 0.05 \\
$K^0_S$           & 73.2 & 62.1 & 41.51 & 27.73  & 16.16& 7.68 \\
$\phi$            & 6.37 & 5.46 & 3.23 & 1.92  & 0.98 & 0.38 \\\hline
\end{tabular}
\end{center}
\caption{Prediction of further hadron multiplicities  using
the parameters as displayed in Figs.~1 - 4.}
\label{predictions}
\end{table}
\newpage
\begin{figure}[tbh] 
\centerline{
\psfig{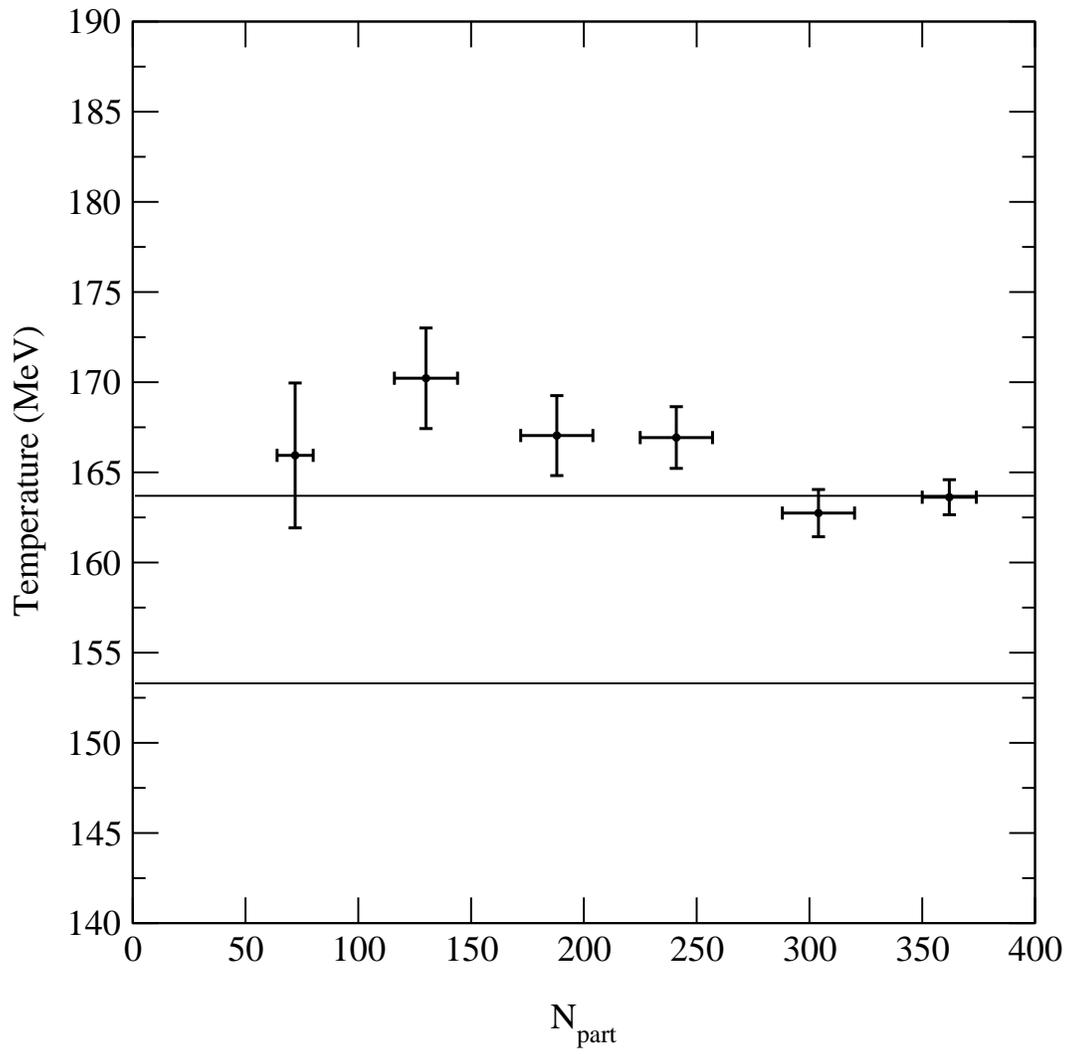}}
\caption{Centrality dependence of the chemical freeze-out temperature
$T$ as a function of the mean participant number. 
The result of \cite{Becattini} is indicated 
by the band: $T = 158.1 \pm 3.2$ MeV.
\label{fig:T}}
\end{figure}

\newpage

\begin{figure}[tbh] 
\centerline{
\psfig{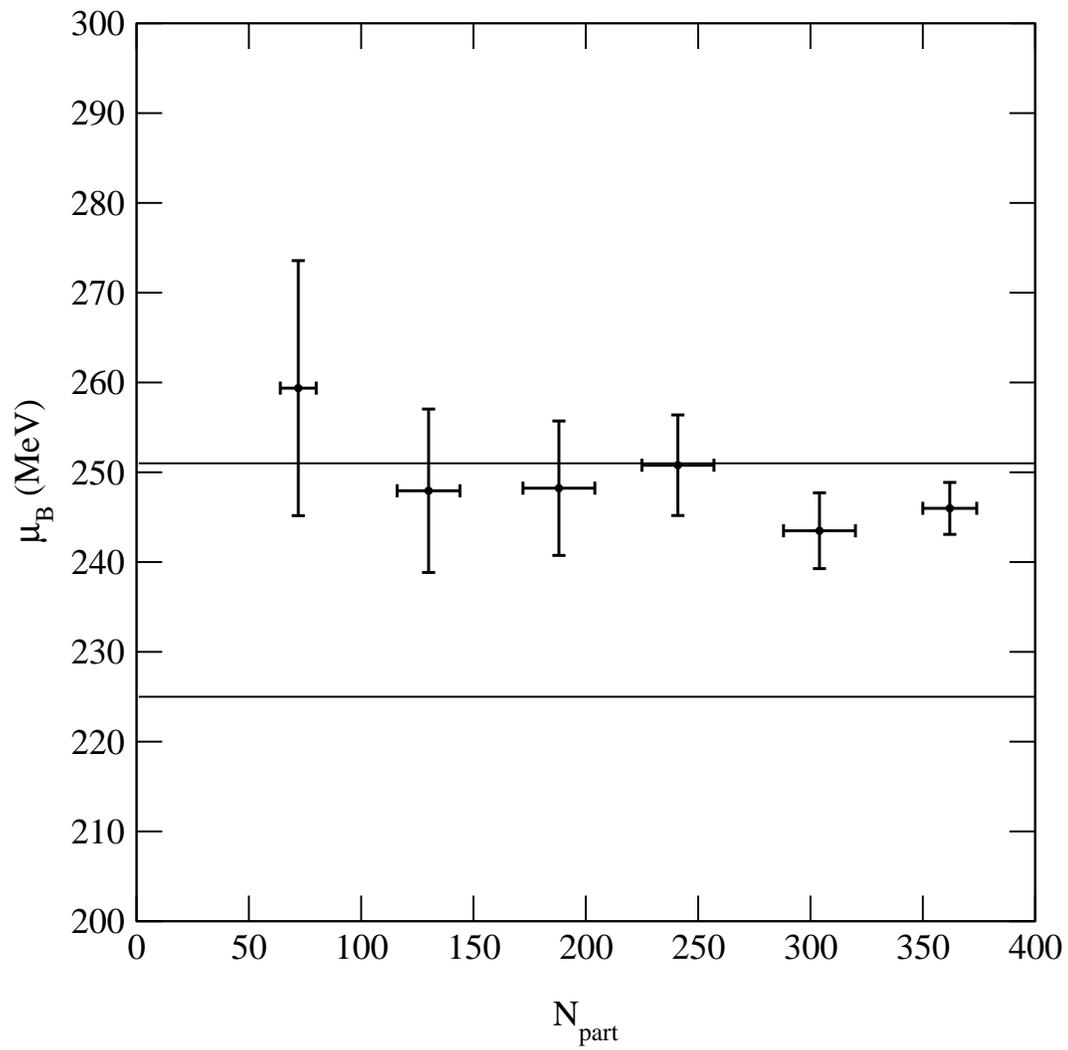}}
\caption{Centrality dependence of the baryon chemical potential $\mu_B$.
The result of \cite{Becattini} is indicated 
by the band: $\mu_B = 238 \pm 13$ MeV.
\label{fig:muB}}
\end{figure}

\newpage

\begin{figure}[tbh] 
\centerline{
\psfig{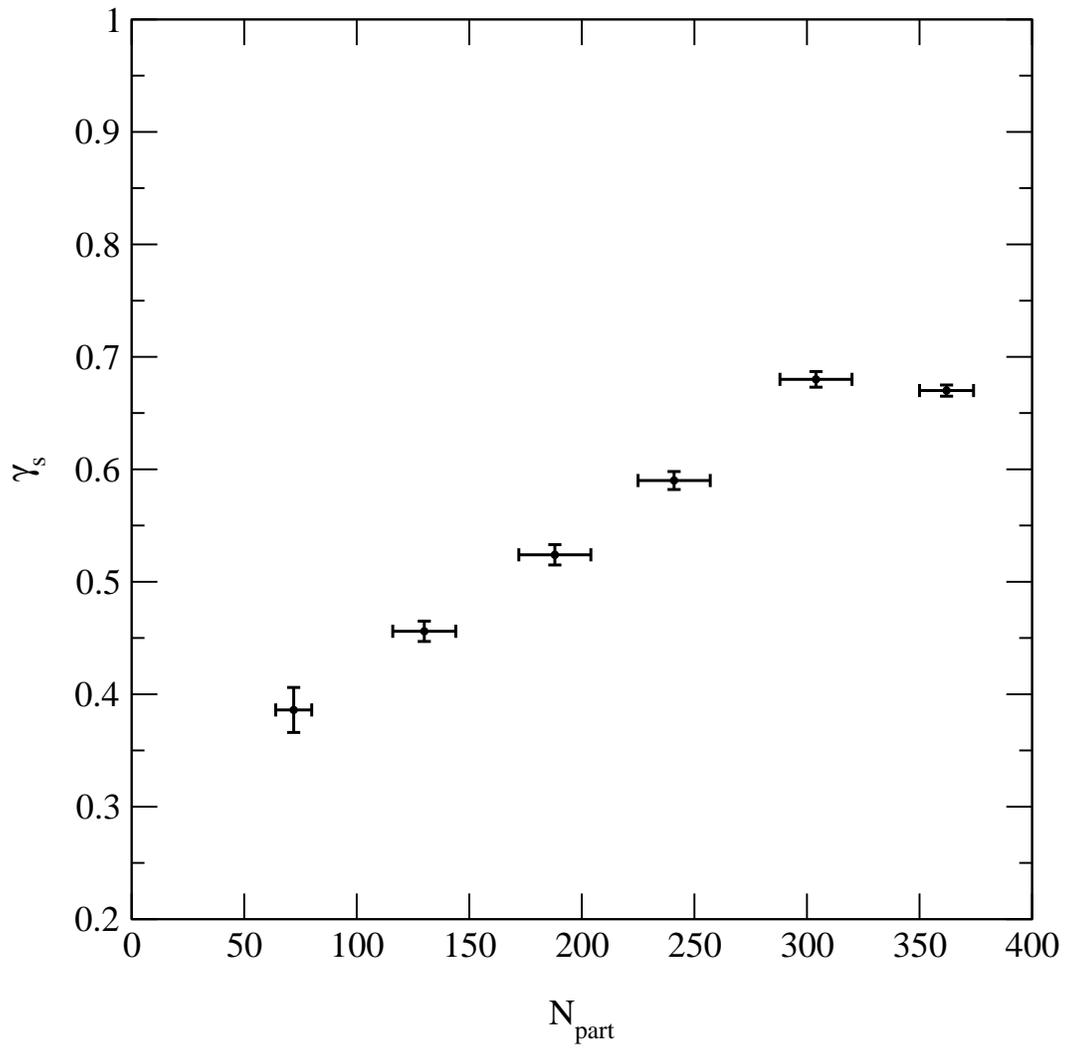}}
\caption{Centrality dependence of the strangeness equilibration 
factor $\gamma_s$. 
\label{fig:gammas}}
\end{figure}

\newpage

\begin{figure}[tbh] 
\centerline{
\psfig{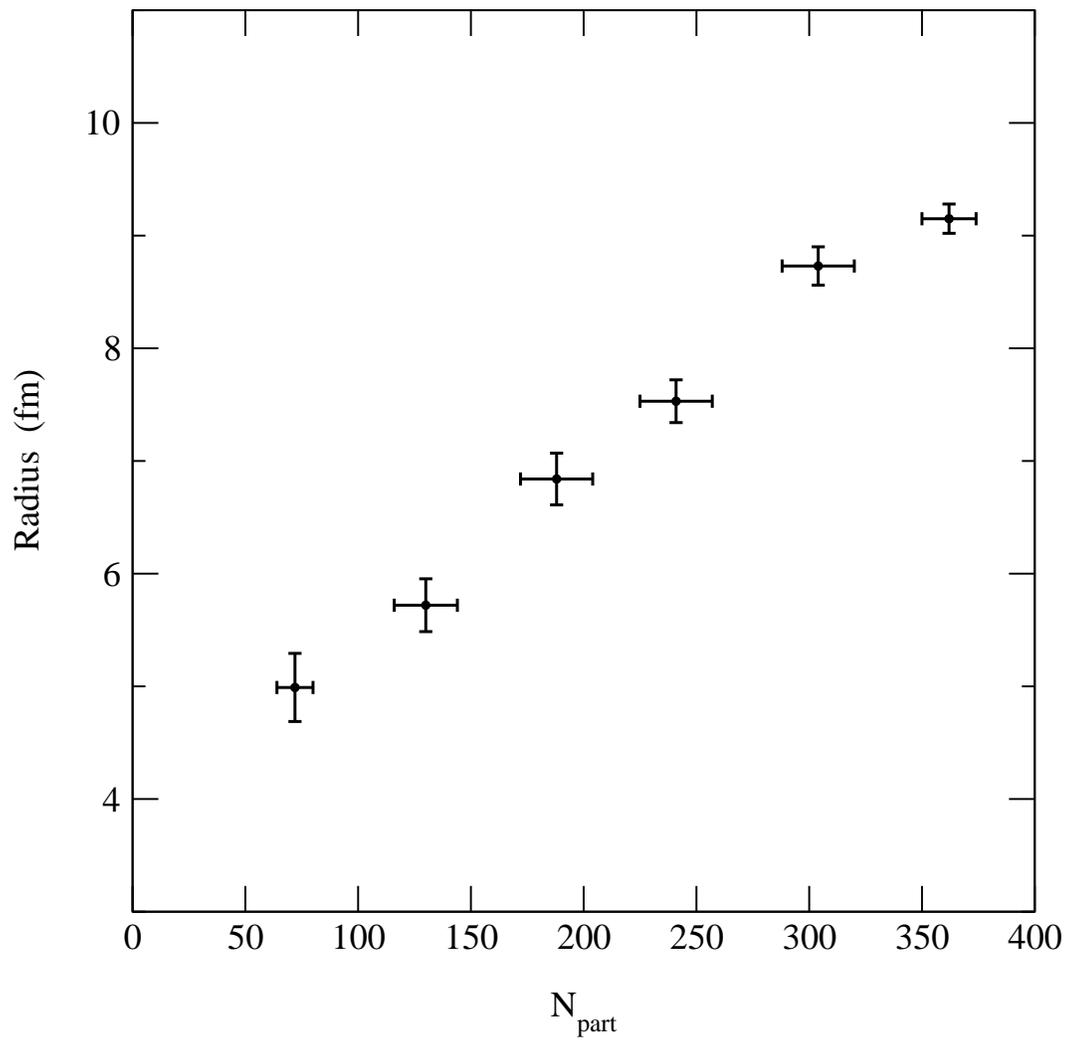}}
\caption{ Centrality dependence of the fireball radius $R$.
\label{fig:R}}
\end{figure}

\newpage

\begin{figure}[tbh] 
\centerline{
\psfig{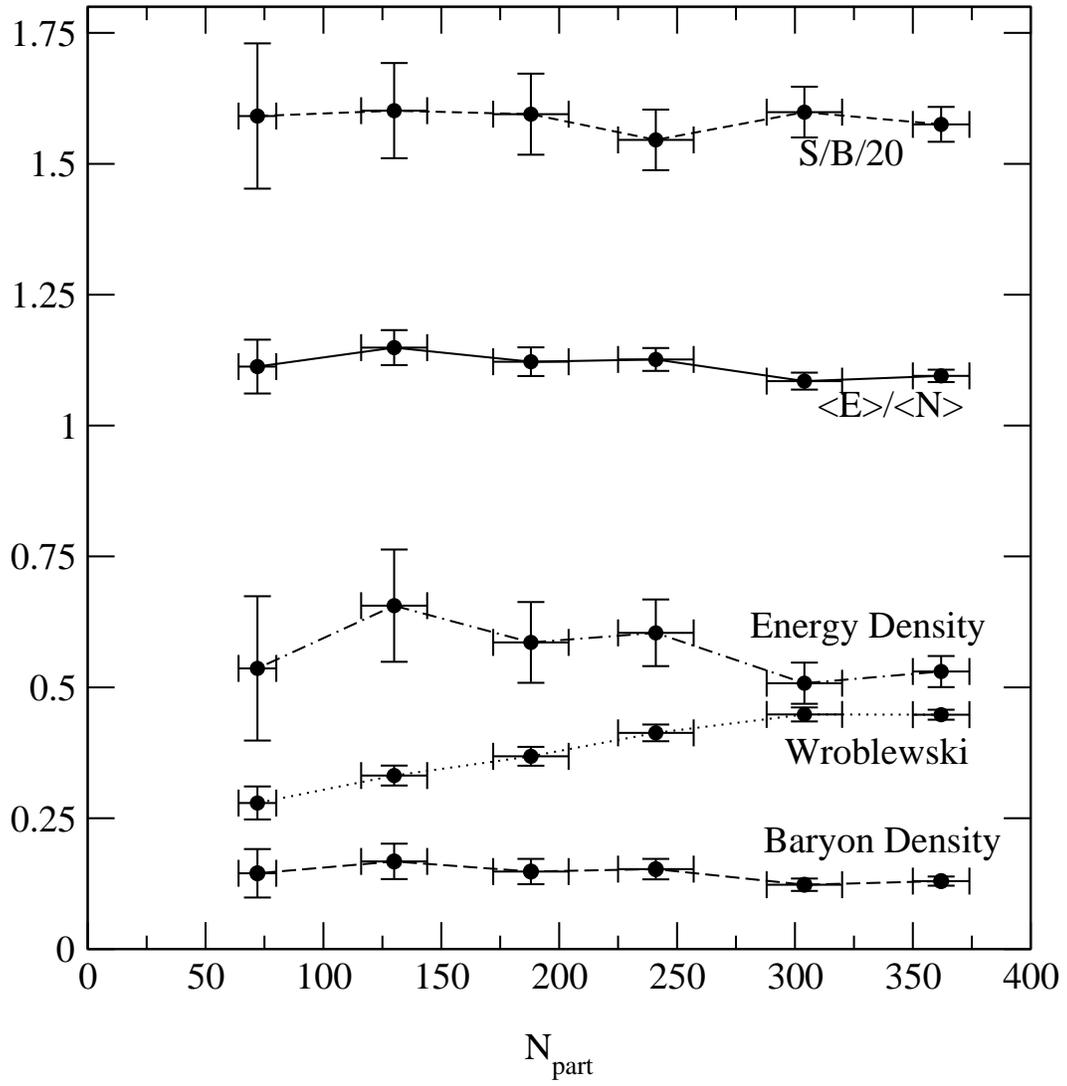}}
\caption{Centrality dependence of various thermodynamic state variables.
The entropy per baryon is scaled by a factor 1/20. The baryon density 
is in units of 1/fm$^3$, the energy density is in units of
 GeV/fm$^3$ and are both uncorrected for excluded volume effects.
The energy per hadron $\left> E\right>/\left< N\right>$ is in units of GeV.
\label{fig:entropy}}
\end{figure}
\end{document}